# Structural and electrical properties of c-axis oriented $Y_{1-x}Ca_xBa_2(Cu_{1-y}Zn_y)_3O_{7-\delta}$ thin films grown by pulsed laser deposition

S.H. Naqib[1,2*], R.A. Chakalov[3], and J.R. Cooper[1]

[1]IRC in Superconductivity, University of Cambridge, Madingley Road, CB3 OHE, UK

[2]MacDiarmid Institute for Advanced Materials and Nanotechnology, Industrial Research Ltd., P.O. Box 31310, Lower Hutt, New Zealand

[3]School of Physics and Astronomy, University of Birmingham, Birmingham B15 2TT, UK

**Abstract**

Ca- and Zn-subsituted $Y_{1-x}Ca_xBa_2(Cu_{1-y}Zn_y)O_{7-\delta}$ (x = 0, 0.05 and y = 0, 0.02, 0.04, 0.05) thin films were grown on $SrTiO_3$ (*100*) substrates using the pulsed laser deposition (PLD) technique. Effects of various growth parameters on the quality of the film were studied via X-ray diffraction (XRD), atomic force microscopy (AFM), and in-plane resistivity, $\rho_{ab}(T)$, measurements. The deposition temperature and oxygen partial pressure were gradually increased to 820°C and 1.20 mbar respectively. Films grown under these conditions exhibited good c-axis orientation (primarily limited by the grain size) and low values of the extrapolated residual resistivity, $\rho(0)$, at zero temperature. The planar hole content, p, was determined from the room temperature thermopower, *S[290K]*, measurements and the effects of oxygen annealing were also studied. Fully oxygenated samples were found to be overdoped with p ~ 0.195. The superconducting transition temperature $T_c(p)$, and $\rho(T,p)$ showed the expected systematic variations with changing Zn content.

# 1. Introduction

YBa$_2$Cu$_3$O$_{7-\delta}$ (Y123) is the most extensively studied compound among all the high-T$_c$ cuprates. But one problem is that with this system, properties of the strongly overdoped (OD) side of the T$_c$-p phase diagram cannot be studied. Pure Y123 with full oxygenation ($\delta = 0$), is only slightly OD (p ~ 0.185, whereas superconductivity is expected to exist up to p ~ 0.27) [1-3]. Further overdoping can only be achieved by substituting Y$^{3+}$ by Ca$^{2+}$, which adds hole carriers to the CuO$_2$ planes of Y123 independently from the oxygenation of the Cu-O chains [1,2]. Recently, we have obtained interesting results regarding the origin, temperature, disorder, and p-dependencies of the pseudogap from the transport and magnetic studies on high-quality sintered polycrystalline samples of Y$_{1-x}$Ca$_x$Ba$_2$(Cu$_{1-y}$Zn$_y$)O$_{7-\delta}$ [3-6]. Although transport and magnetic studies on single crystals give more information, it is difficult to grow homogeneous Ca-Y123 single crystals.

* Corresponding author: Phone: +64-4-9313676; Fax: +64-4-9313117; E-mail: s.naqib@irl.cri.nz

Also successful growth of Y$_{1-x}$Ca$_x$Ba$_2$(Cu$_{1-y}$Zn$_y$)O$_{7-\delta}$ single crystals has not been reported. Ca-Zn-Y123 is an important system for understanding the fundamental physics of cuprates [3-6]. Epitaxial thin films of Ca-substituted or Zn-substituted Y123 have been grown by several groups [7-10] but, so far, films containing both Zn and Ca have not been fabricated, to our knowledge. This is also potentially an important system for practical purposes as Ca- and Zn-substituted overdoped Y123 can have a high critical current density due to highly conducting grain boundaries produced by Ca [11] and pinning of vortices by Zn [12].

In this paper, we report the systematic changes in the pulsed laser deposition (PLD) process, needed to grow high-quality c-axis oriented Y$_{1-x}$Ca$_x$Ba$_2$(Cu$_{1-y}$Zn$_y$)$_3$O$_{7-\delta}$ thin films. We also report the results of X-ray diffraction (XRD), oxygen annealing, room temperature thermopower (*S[290K]*), atomic force microscopy (AFM), and resistivity measurements used to characterize these films. It turns out that with the notable exception of T$_c$(p) and *S[290K]*, the other measured properties are mainly determined by the average grain size of the films.

## 2. Deposition of the film:

The high density (> 90% of the ideal X-ray density) phase pure $Y_{0.95}Ca_{0.05}Ba_2(Cu_{1-y}Zn_y)_3O_{7-\delta}$ targets were prepared using standard solid state synthesis [3]. A commercially available target (supplied by the company *PiKem*) was used for $YBa_2Cu_3O_{7-\delta}$ films. All films were grown on 10 x 5 x 1mm$^3$ SrTiO$_3$ substrates, highly polished on one side. The crystallographic orientation of the substrate was (100) (SrTiO$_3$ is a cubic system with a lattice constant of 3.905 Å) [13]. The films were fabricated in Birmingham using a *Lambda Physik* LPX210 KrF laser with a wavelength of 248 nm. The process of PLD works in the following way [14] - when the laser radiation is absorbed by the target surface, electromagnetic energy is converted into thermal, chemical, and even mechanical energy to cause rapid evaporation, ablation, excitation, plasma formation, and exfoliation. Evaporants form a plume consisting of a mixture of energetic species including atoms, molecules, electrons, ions, clusters, micron-sized solid particulates, and molten globules. The ability of the PLD technique to transfer complex target compositions to a substrate with relative ease under the appropriate conditions is one of the key features of this process.

The laser intensity inside the chamber was measured before each PLD run and was adjusted to the desired value using focussing/attenuating systems. We have always used a pulse rate of 10 Hz during thin film deposition. Prior to the deposition, the substrate heater block was fully out-gassed by heating it to the deposition temperature with the turbo pump running until the chamber pressure reached ~ $10^{-5}$ mbar. A pre-ablation run was always performed using 1000 laser pulses at 10 Hz (same frequency was used during ablation) to clean the rotating target. The deposition temperature, $T_{ds}$, is the temperature of the heater block measured by a thermocouple fixed to it. The selected target-substrate distance depended on the oxygen partial pressure, $PO_2$, and on the incident laser energy, as these two were the main parameters determining the size and shape of the plume. Relevant information regarding the deposition conditions and films studied are given in Table 1.

## 3. Results:

### A. X-ray:

XRD measurements were performed with a *Philips* PW1050 vertical goniometer head mounted on a PW1730 X-ray generator and controlled by *Philips* X'pert software. The X-ray (intensity versus 2θ [in degree]) results for various films are shown in Figs.1. Results from rocking-curve ($\Omega$-scan) analysis are shown in the insets. These are often interpreted as measuring the degree of c-axis orientation of the films and this is briefly discussed later. We have used the (007) diffraction peak for rocking-curve analysis, following [13]. The expected X-ray diffraction pattern from a c-axis oriented film would be a series of (00$l$) reflections. A very clear and sharp set of (00$l$) peaks were obtained for our PLD films with no sign of any impurity peak (see Figs.1). With a $SrTiO_3$ substrate, the situation is slightly complicated by the fact that the c-axis spacing is very close to one-third of the Y123 (and $Y_{0.95}Ca_{0.05}Ba_2(Cu_{1-y}Zn_y)_3O_{7-\delta}$) c-axis spacing. This leads to a diffraction pattern in which every third peak - those labeled (003$n$) with $n$ integer - is obscured by an intense reflection from the substrate. One of the remaining peaks can be used for rocking-curve analysis and used as a basis for comparing the degree of c-axis orientation for different samples. Most of our films show a narrow (007) peak indicative of well-oriented crystalline state. These peaks have double structures, arising from the $CuK_{\alpha 2}$ component of the incident X-rays. This double structure becomes more evident as $T_{ds}$ is increased and the profiles of the rocking-curves become sharper, as shown in Fig.2a. The results from Fig.2a are summarized in Table 2a. We have also examined the effect of oxygen partial pressure, Fig.2b shows the profiles of the (007) peaks of $Y_{0.95}Ca_{0.05}Ba_2Cu_3O_{7-\delta}$ films grown with different oxygen partial pressures. We have summarized the results shown in Fig. 2b in Table 2b. It is clear from Tables 2a and 2b that high deposition temperature and high oxygen partial pressure improve the film quality, as determined both by the angular widths and peak heights of the rocking curves. We note that the thickness of all these films was the same within ± 8% and XRD was done under the same conditions.

**B. AFM results:**

AFM was employed to determine the thickness and to estimate the grain sizes of the films. The thickness of the films lies within the range (3000 ± 200) Å. The average grain sizes determined by AFM for four typical films B1, B4, B5, and B9 were 0.34, 0.22, 0.15, and 0.32 ± 0.05 μm respectively and increased systematically with the deposition temperature $T_{ds}$. There were only a few out-growths but boulders of other phases were always present, typically covering less than 3% of the surface area of the film.

**C. Oxygenation, *S[290K]*, and essential resistive features:**

The oxygenation conditions, the results of room-temperature thermopower (*S[290K]*) measurements, and the resistive features directly associated with the quality of the films are presented in Table 3. It is clearly seen from Table 3 that the best thin-film samples are grown at deposition temperatures in the range 800°C to 820°C with oxygen pressures in the range 0.95 to 1.20 mbar.

The best samples, for a given Zn content, are resistively characterized by low values of ρ(300K) and R(0K)/R(300K), where R(0K) is the resistance extrapolated to zero temperature from above $T_c$.

For a given starting composition, the $T_c$ values of the different quality samples did not change significantly, so based on the work of Rullier-Albenque *et al.* [15], we believe that the residual resistivity is mainly determined by grain boundary resistance for our thin films. In ref.[15] it is suggested that in a d-wave superconductor, there is a direct link between intragrain residual resistivity and the value of $T_c$. Therefore, if $T_c$ is high the apparent residual resistivity must arise from grain boundaries. It can be also seen from the $T_c$-values in Table 3 that the level of substitution of Zn in place of in-plane Cu becomes harder for Zn concentrations above 4% (y) in $Y_{0.95}Ca_{0.05}Ba_2(Cu_{1-y}Zn_y)_3O_{7-\delta}$.

The resistive features of our best thin film samples are comparable to the best reported in the literature [7-10,12,16-18]. The important effect of high-temperature oxygenation at $T_{ds}$, is clearly seen from Table 3. Oxygenation at these elevated temperatures always reduces *S[290K]*, which varies systematically with p over a wide range for different families of cuprate superconductors [19] and decreases with increasing p. It is also independent of grain boundary resistance and isovalent atomic disorder such as Zn substitution. Reliable estimates of p can be obtained from *S[290K]* using the equations [19]:

*S[290K]* = -139p +24.2    for p > 0.155

*S[290K]* = 992exp(-38.1p) for p ≤ 0.155

From the room-temperature thermopower measurements our most overdoped sample (B7) has a planar carrier concentration of p = 0.195 ± 0.004 and $T_c$ = 82.6 ± 1 K. This sample has extremely low values of ρ(300K) and R(0K)/R(300K), and a very high degree of crystalline c-axis orientation (Table 2a). Therefore, we believe that it represents the intrinsic ab-plane behavior to a large extent.

**D. Resistivity:**

The in-plane resistivity, $\rho_{ab}(T)$, for different Zn and hole contents are presented in this section. Details of these measurements can be found in Ref. [3]. $\rho_{ab}(T)$ of nearly optimally doped (p ~ 0.157) pure Y123 is shown in Fig.3a. The high quality of the film is evident from the low values of ρ(300K) as well as the extrapolated R(0K)/R(300K). The extremely narrow resistive (10% to 90%) transition width of ~ 0.40K is indicative of high degree of homogeneity in this film.

We have located the pseudogap temperature, $T^*(p)$, as the temperature at which experimental ρ(T) starts falling below the high-T linear fit to the resistivity data, given by $\rho_{LF}$ (see the inset of Fig.3a). This decrease in the resistivity is attributed due to a suppressed electronic density of states (EDOS) near the Fermi level [20] and $T^*(p)$ should be interpreted as the temperature at which this effect from the suppressed EDOS becomes detectable in ρ(T) [21]. The value of the characteristic pseudogap temperature, $T^*$ (~ 125 ± 5K), of this optimally doped Y123 thin film agrees quite well with those found for other sintered polycrystalline $Y_{1-x}Ca_xBa_2(Cu_{1-y}Zn_y)_3O_{7-\delta}$ samples studied earlier [3-5] for the same hole concentration, emphasizing the fact that $T^*(p)$ is fairly insensitive to the crystalline state and disorder contents of the compounds. We have shown $\rho_{ab}(T)$ for another nearly optimally doped (p = 0.154 ± 0.004) $Y_{0.95}Ca_{0.05}Ba_2Cu_3O_{7-\delta}$ film in Fig.3b. For comparison, $\rho(T)$ for sintered $Y_{0.95}Ca_{0.05}Ba_2Cu_3O_{7-\delta}$ with similar hole content (p = 0.151 ± 0.004) has also been included in this figure. The two data sets are plotted on different scales so that the slopes, $d\rho(T)/dT$, appear to be the same. The difference in magnitudes of $d\rho(T)/dT$ (by a factor 3.4) arises from the porosity of the sintered sample [22] which causes the current to flows percolatively following the ab-plane paths rather than the higher resistance c-axis ones [22]. The slightly higher residual resistivity of the sintered sample (relative to $d\rho(T)/dT$) is probably a grain boundary effect. Even though the *S[290K]* data suggest that the two values of p are same within the experimental error, the film has a slightly lower $T_c$ than the ceramic, 86.4 and 88.1 respectively. We have found $T_c$ to be consistently lowered by ~ 2K for the films compared with the $T_c$ for sintered samples at the same hole concentration.

A possible reason for this might be the non-uniform epitaxial strain due to the lattice mismatch between the target material and the substrate. Alternatively it could be caused by in-plane atomic disorder in the films. According to the proposed universal relation between $T_c$ and $\rho(0)$ for cuprate superconductors [15] (a d-wave effect whereby intragrain scattering mixes positive and negative components of the order parameter), a shift of 2K in $T_c$ corresponds to an intragrain residual resitivity of only 4 $\mu\Omega$ cm, and in our work this is not detectable relative to the larger grain boundary contribution. The resistive transition widths (taken as the difference between 90% and 10% resistivity points) for these two samples are 1.0K (film) and 1.4K (sintered).

Effects of Zn substitution on $\rho_{ab}(T)$ for the films with similar hole concentrations (in the range p = 0.162 ± 0.010) have been illustrated in Fig.3c. The residual resistivity, $\rho(0)$, increases linearly with increasing Zn, as shown in Fig. 3d. Matthiessen's rule is obeyed in these films in the sense that slopes of the $\rho_{ab}(T)$ data are fairly insensitive to Zn contents (Fig. 3c). All these results are in quantitative agreement with those found for sintered polycrystalline samples [3] and earlier studies on Zn-substituted Y123 thin films [8]. Another important parameter is the rate of suppression of $T_c$ with Zn. Fig.4 shows $T_c(y)$ for the films shown in Fig. 3c. The value of $dT_c/dy$ (= -9.6 K/%Zn) for these films agrees quite well with those obtained for sintered polycrystalline samples with similar hole contents [3].

Representative plots of $\rho_{ab}(T)$ data for $Y_{0.95}Ca_{0.05}Ba_2Cu_3O_{7-\delta}$ thin films with different growth parameters are shown in Fig.3e. This figure clearly illustrates the significant roles played by the deposition conditions summarized in Table 1. Invariably films (at a given p) with superior resistive features (like, low values of $\rho(300K)$, $\rho(0)$, and superconducting transition width) are those which also show higher degree of crystalline orientation (sharp (007) peaks) and larger grains.

### 4. Discussions and conclusion

We know from previous work [19] that the room-temperature thermoelectric power is not affected by grain boundary or isovalent atomic substitutions (e.g., Zn for in-plane Cu) and gives a reliable measure of p [3,19]. In the present work *S[290K]* varies from +7.0 to -3.0 µV/K, corresponding to p-values between 0.132 and 0.195, depending on the deposition conditions and *in-situ* oxygen annealing. The resistivity curves (Fig.3a and Fig.3e) correlate well with the sizes of the grains and the sharpness of the rocking curves. We believe that the main effects on $d\rho(T)/dT$ are (a) the doping level (p-values) and (b) possible contribution from enhanced electrical conductivity of the Cu-O chains in the better samples. However the main effect on $\rho(0)$ is the size of the grains and the fact that grain boundary resistances are more important for smaller grain sizes.

With regard to the X-ray rocking curves - firstly, there is a clear correlation between the sizes of the grains and the absolute peak intensity. This is expected theoretically, as the peak intensity should scale as the square of the grain size (square of the number of atoms scattering coherently). There is also a correlation between the angular width (FWHM) of the rocking curves and the grain size. We have not been able to account for this correlation in terms of a model calculation in which the width of the diffraction peak is limited by the number of unit cells in a grain (the observed broadening of the rocking curves is a factor of 3 too large). Instead, we believe that it arises from the spiral growth of the films [23]. In support of this we note that a typical variation of one c-axis lattice parameter (~ 11.8Å) across a grain size of 0.22μm (for B4) corresponds to c-axis being tilted by $0.30^o$. This is in good agreement with the FWHM of $0.28^o$ for this particular film (Table 2a).

In conclusion we have correlated the structural and transport properties of epitaxial $Y_{1-x}Ca_xBa_2(Cu_{1-y}Zn_y)_3O_{7-\delta}$ while varying their growth parameters.

Detailed measurements of resistivity and magneto-resistivity for these films are in progress over a wide range of hole concentrations, which we believe can yield valuable information regarding the nature of the pseudogap.

**Acknowledgements:**

We are grateful to Dr. A. Kursumović for taking the AFM pictures that allowed estimation of the grain size and Dr. C. Muirhead for making the PLD facilities in Birmingham University available. We also thank Mr. J. Ransley for his help with other AFM measurements. SHN acknowledges financial support from the Commonwealth Scholarship Commission (UK), Darwin College, Department of Physics, and the IRC in Superconductivity.

**Figure captions:**
(S.H. Naqib *et al.*: Structural and electrical……..)

Fig.1. X-ray diffraction profiles for (a) B6, (b) B11, and (c) B16 (see Table 1 for film compositions).

Fig.2. (a) Effect of the deposition temperature $T_{ds}$, on the profile of (007) peak for $Y_{0.95}Ca_{0.05}Ba_2Cu_3O_{7-\delta}$ films. All these films were grown at an oxygen partial pressure of 0.95 mbar. (b) Effect of the oxygen partial pressure $PO_2$, on the profile of (007) peak for $Y_{0.95}Ca_{0.05}Ba_2Cu_3O_{7-\delta}$ films. All these films were grown at a deposition temperature of 800°C.

Fig.3. (a) Main: $\rho_{ab}(T)$, $\rho_{LF}$, and the residual resistivity $\rho(0)$, for optimally doped Y123 (B1). Insets: (top) detection of $T^*$ (see text); (bottom) R(T)/R(300K), the dashed straight line gives the extrapolated R(0)/R(300K).
(b) Resistivity of sintered and c-axis thin film (B6) of $Y_{0.95}Ca_{0.05}Ba_2Cu_3O_{7-\delta}$. Both these samples are slightly underdoped.
(c) Effect of Zn on $\rho_{ab}(T)$ for 5%Ca substituted thin films. Dashed straight lines are linear fits near the superconducting transition used to obtain the values of $\rho(0)$.
(d) $\rho(0)$ vs. Zn (y in %) for $Y_{0.95}C_{0.05}Ba_2(Cu_{1-y}Zn_y)_3O_{7-\delta}$ thin films. The dashed straight line is drawn as a guide to the eye.
(e) $\rho_{ab}(T)$ for $Y_{0.95}Ca_{0.05}Ba_2Cu_3O_{7-\delta}$ thin films with different growth parameters.

Fig.4. $T_c$(y in %) for $Y_{0.95}C_{0.05}Ba_2(Cu_{1-y}Zn_y)_3O_{7-\delta}$ thin films. Linear fit to $T_c(y)$ is shown by the dashed line. The fitting equation is also shown.

Table 1: **Deposition conditions and film identities (Film-ID).**

| Film-ID (composition) | $T_{ds}$(°C) | $PO_2$ (mbar) | Target-substrate distance (cm) | Number of pulses | Energy density on target (J/cm$^2$) |
|---|---|---|---|---|---|
| B1 Y123 | 760 | 0.95 | 6.5 | 2000 | 1.5 |
| B2 (5%Ca-0%Zn) | 780 | 0.95 | 6.5 | 2000 | 1.5 |
| [B3 (5%Ca-0%Zn); all parameters identical to B2, but different *in situ* oxygenation conditions] | | | | | |
| B4 (5%Ca-0%Zn) | 740 | 0.95 | 6.5 | 2000 | 1.5 |
| [B5 (5%Ca-0%Zn); all parameters identical to B4, but different *in situ* oxygenation conditions] | | | | | |
| B6 (5%Ca-0%Zn) | 800 | 0.95 | 6.5 | 2000 | 1.5 |
| B7 (5%Ca-0%Zn) | 820 | 0.95 | 6.5 | 2000 | 1.5 |
| B8 (5%Ca-0%Zn) | 800 | 0.70 | 6.5 | 2000 | 1.5 |
| B9 (5%Ca-0%Zn) | 800 | 1.20 | 6.5 | 2000 | 1.5 |
| B10 (5%Ca-2%Zn) | 800 | 0.95 | 6.5 | 2000 | 1.5 |
| B11 (5%Ca-2%Zn) | 820 | 0.95 | 6.5 | 2000 | 1.5 |
| B12 (5%Ca-4%Zn) | 800 | 0.95 | 6.5 | 2000 | 1.5 |
| B13 (5%Ca-4%Zn) | 820 | 0.95 | 6.5 | 2000 | 1.5 |
| B14 (5%Ca-5%Zn) | 800 | 0.95 | 6.5 | 2000 | 1.5 |
| B15 (5%Ca-5%Zn) | 820 | 0.95 | 6.5 | 2000 | 1.5 |
| B16 (5%Ca-5%Zn) | 820 | 1.20 | 6.5 | 2000 | 1.5 |

Table 2a: **Full-widths at half-maximum of (007) peaks shown in Fig.2a.**

| Film | $T_{ds}$(°C) | FWHM of (007) peak (in degree) |
|---|---|---|
| B4 | 740 | 0.280 |
| B6 | 800 | 0.184 |
| B7 | 820 | 0.156 |

Table 2b: **Full-widths at half-maximum of (007) peaks shown in Fig.2b.**

| Film | $PO_2$(mbar) | FWHM of (007) peak (in degree) |
|---|---|---|
| B8 | 0.70 | 0.340 |
| B6 | 0.95 | 0.180 |
| B9 | 1.20 | 0.121 |

Table 3: **Oxygenation conditions, *S[290K]*, and some resistive features of the films.**

| Film ID | $S[290K]$ (μV/K) | *R(0K)/R(300K) | ρ(300K) (μΩ cm) | **Oxygenation conditions | ***$T_c$ (K) |
|---|---|---|---|---|---|
| B1 (Y123) | + 2.11 | 0.06 | 218 | Held at $T_{ds}$ for 15 mins, cooled to 450°C @30°C/min and held for 30 mins, then cooled @30°C/min to room-temperature | 91.1 |
| B2 (5%Ca-0%Zn) | + 4.14 | 0.23 | 306 | Held at $T_{ds}$ for 15 mins, cooled to 475°C @30°C/min and held for 30 mins, then cooled @30°C/min to room-temperature | 85.8 |
| B3 (5%Ca-0%Zn) | + 5.06 | 0.28 | 383 | Cooled from $T_{ds}$ without holding, otherwise same as that employed to B2 | 85.2 |
| B4 (5%Ca-0%Zn) | + 6.09 | 0.26 | 518 | Held at $T_{ds}$ for 15 mins, cooling to 500°C @30°C/min and holding for 30 mins, then to room-temperature @ 30°C/min | 84.4 |
| B5 (5%Ca-0%Zn) | + 7.0 | 0.296 | 526 | Cooled from $T_{ds}$ without holding, otherwise same as that employed to B4 | 84.0 |
| B6 (5%Ca-0%Zn) | + 2.91 | 0.15 | 341 | Held at $T_{ds}$ for 15 mins, cooled to 425°C @30°C/min and holding for 30 mins, then cooled to room-temperature @ 30°C/min | 86.4 |
| B7 (5%Ca-0%Zn) | - 3.00 | 0.11 | 184 | Held at $T_{ds}$ for 15 mins, cooled to 350°C @30°C/min, held for 30 mins, cooled to room-temperature @30°C/min | 82.6 |
| B8 (5%Ca-0%Zn) | + 2.81 | 0.21 | 424 | Same as that employed to B6 | 82.9 |
| B9 (5%Ca-0%Zn) | + 3.20 | 0.18 | 294 | Cooled from $T_{ds}$ to 425°C @30°C/min, held for 30 mins, cooled @30°C/min to room-temperature | 84.9 |
| B10 (5%Ca-2%Zn) | - 2.21 | 0.14 | 263 | Same as that employed to B7 | 64.1 |
| B11 (5%Ca-2%Zn) | - 0.06 | 0.183 | 341 | Cooled from $T_{ds}$ without holding, otherwise same as that employed to B7 | 66.16 |
| B12 (5%Ca-4%Zn) | + 0.66 | 0.23 | 401 | Same as that employed to B7 | 47.16 |
| B13 (5%Ca-4%Zn) | + 4.59 | 0.21 | 511 | Cooled from $T_{ds}$ to 450°C @30°C/min, held for 30 mins, cooled @30°C/min to room-temperature | 41.8 |
| B14 (5%Ca-5%Zn) | + 1.93 | 0.22 | 609 | Same as that employed to B7 | 49.95 |
| B15 (5%Ca-5%Zn) | + 1.72 | 0.31 | 581 | Same as that employed to B7 | 39.88 |
| B16 (5%Ca-5%Zn) | + 1.33 | 0.24 | 436 | Same as that employed to B7 | 38.1 |

* R(0K) was obtained from the linear extrapolation of the R(T) data from high temperatures above $T^*$, the pseudogap temperature (see section 3D).
** *In situ* oxygenation was done with an oxygen pressure of 1 ATM for all the films.
*** $T_c$ was taken at $R(T_c) = 0$ Ω, within the noise level ($\pm 10^{-6}$ Ω) of measurement.

Fig.1. (S.H. Naqib *et al.*: Structural and electrical …………..)

(a)
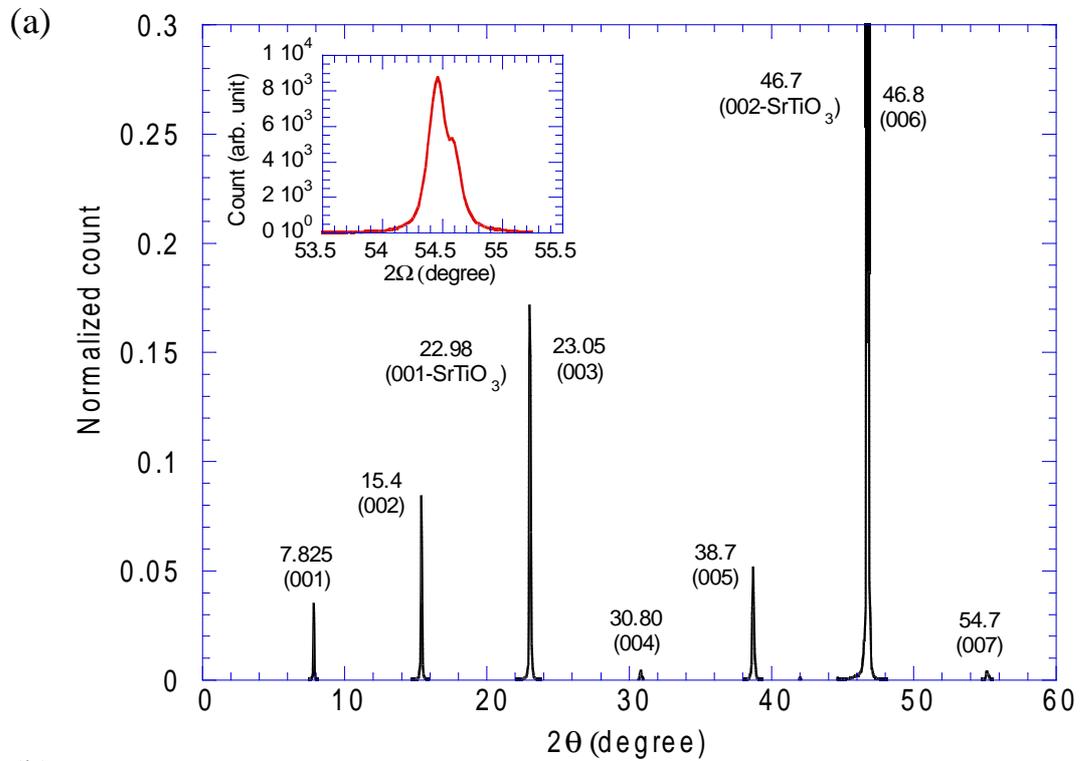

(b)
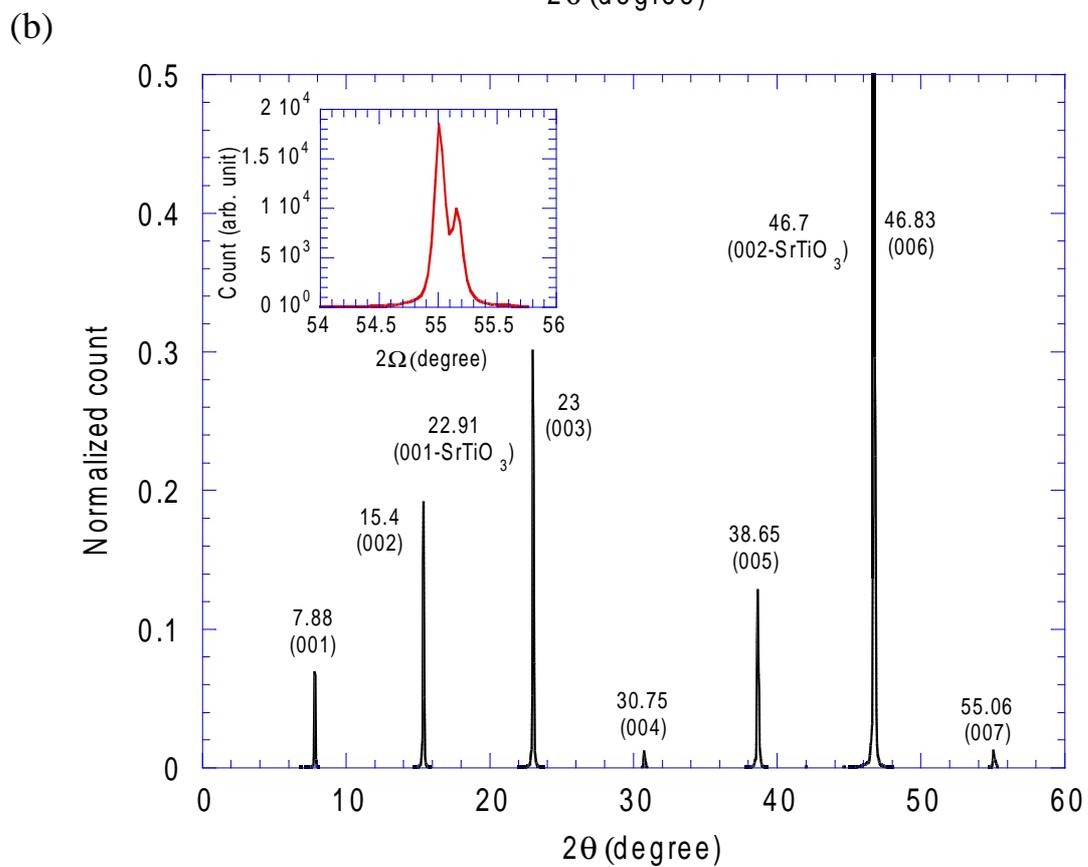

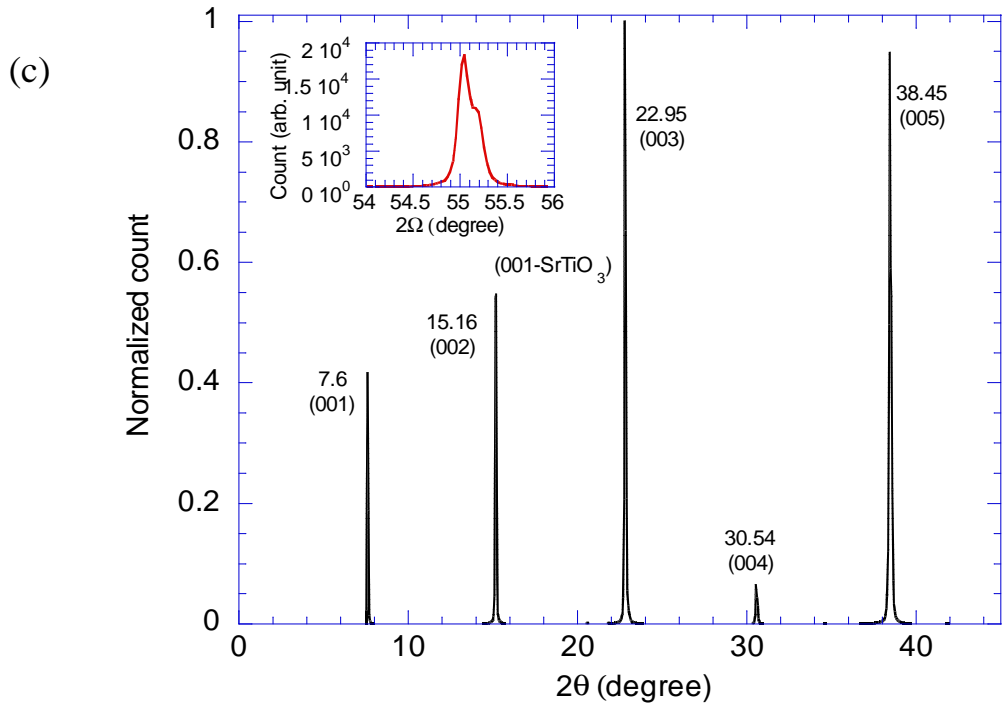

Fig.2. (S.H. Naqib *et al.*: Structural and electrical …………..)

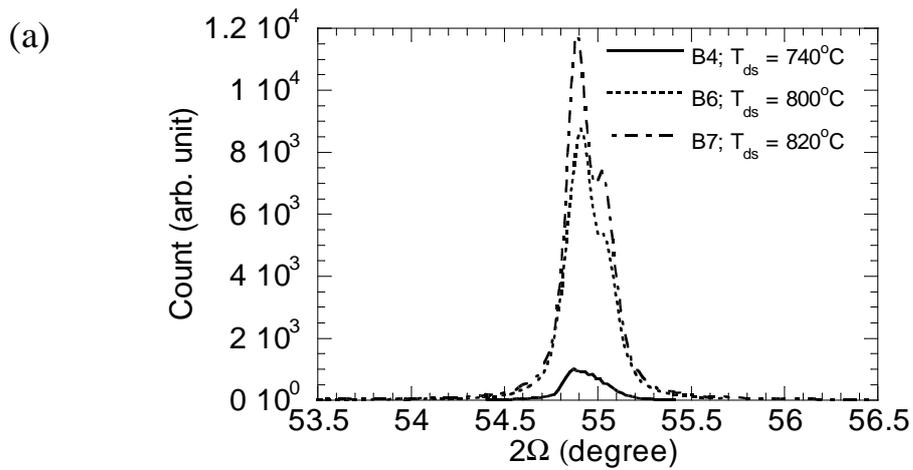

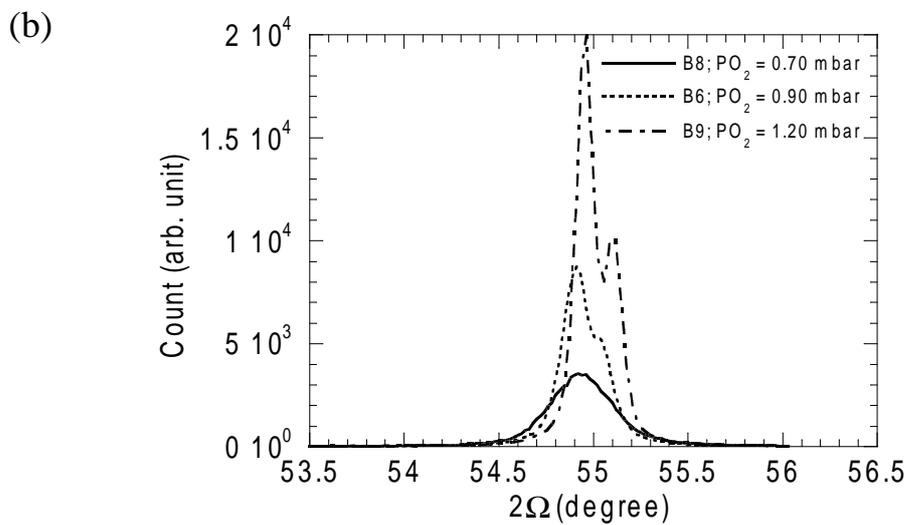

Fig.3. (S.H. Naqib *et al.*: Structural and electrical…………..)

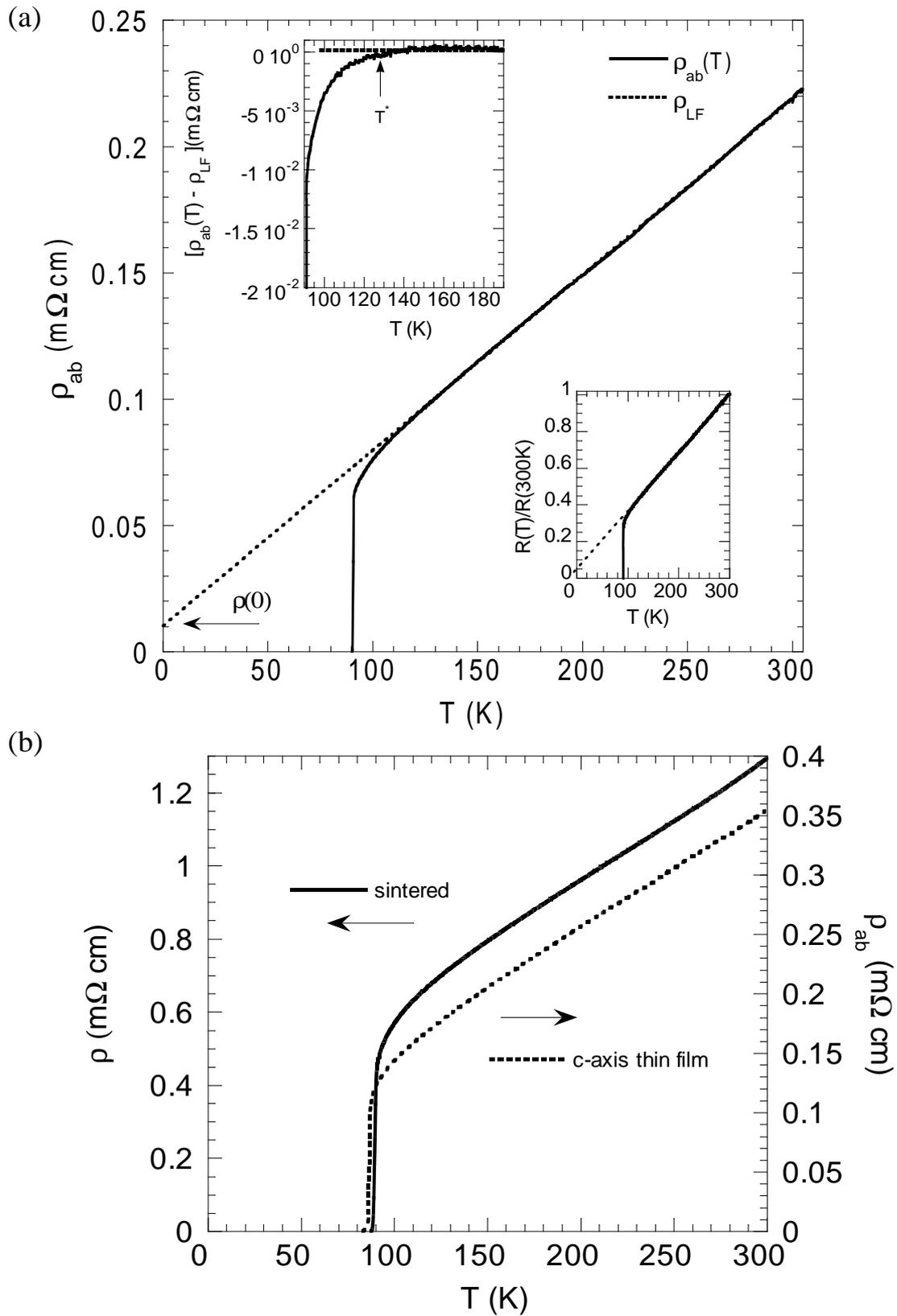

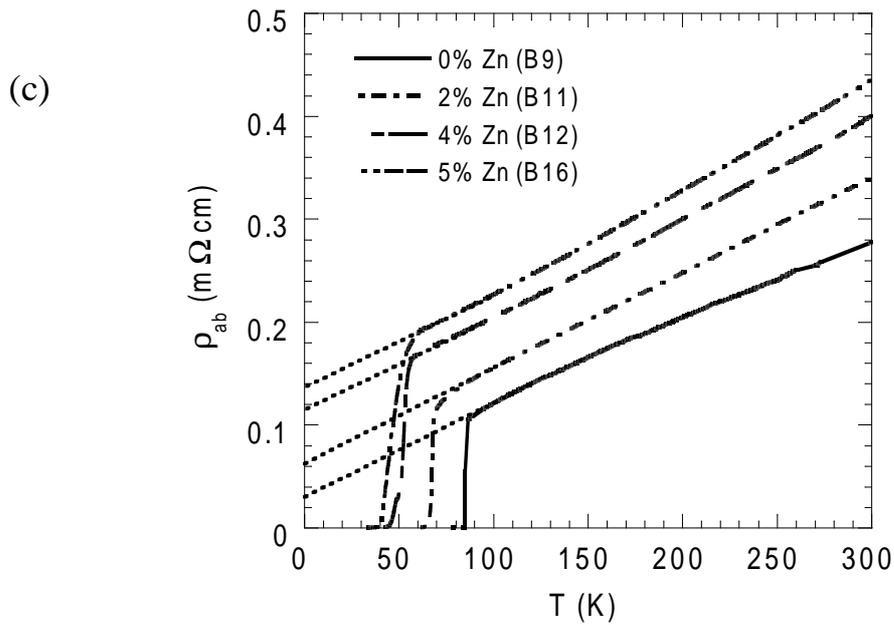

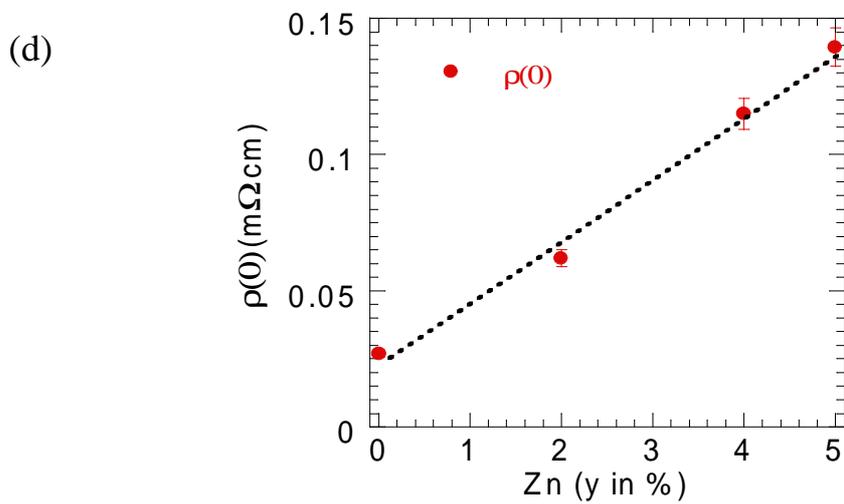

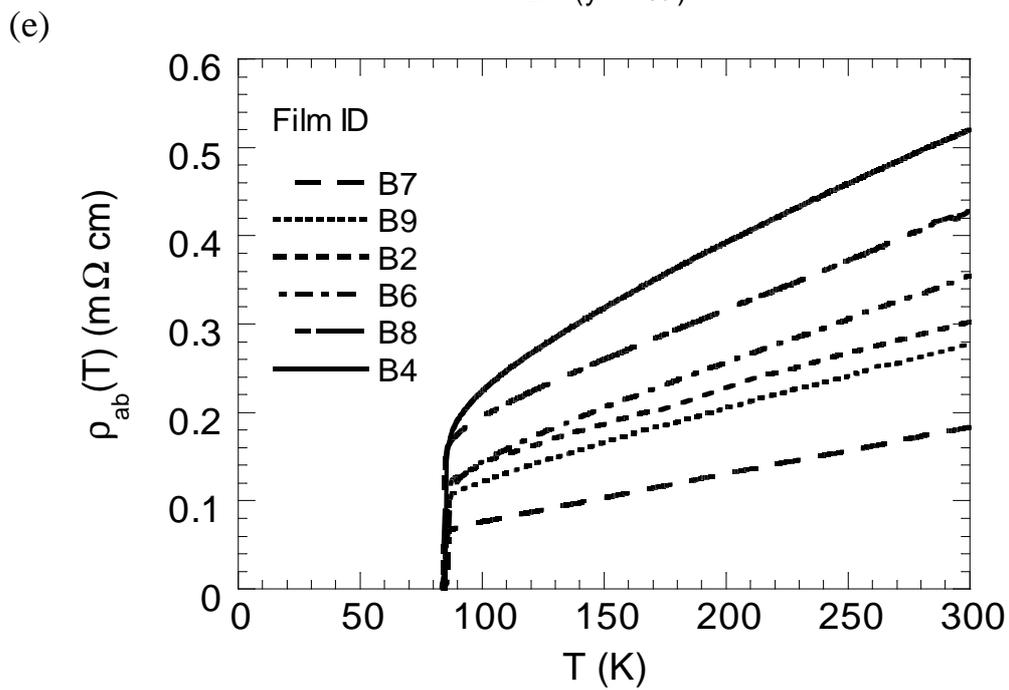

Fig.4. (S.H. Naqib *et al.*: Structural and electrical…………..)

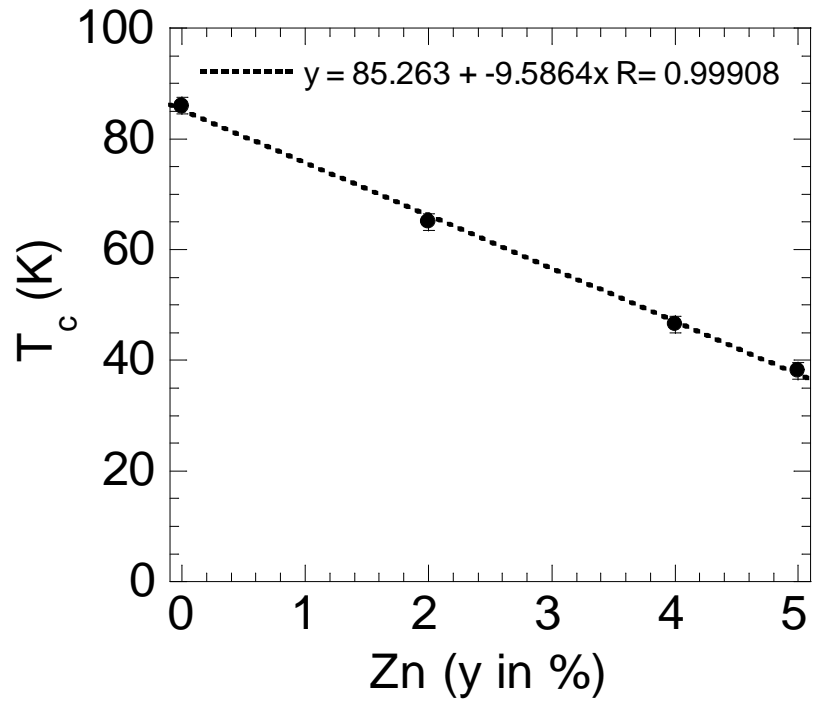